\documentclass[12pt]{article}
\usepackage{graphicx, amsmath, amssymb, cite, setspace, color,fancyheadings}
 \usepackage[top=1 in, bottom=1 in, left=1 in, right=1 in]{geometry}

\newcommand{\captionfonts}{\footnotesize} 
\makeatletter 
\long\def\@makecaption#1#2{%
  \vskip\abovecaptionskip
  \sbox\@tempboxa{{\captionfonts #1: #2}}%
  \ifdim \wd\@tempboxa >\hsize
    {\captionfonts #1: #2\par}
  \else
    \hbox to\hsize{\hfil\box\@tempboxa\hfil}%
  \fi
  \vskip\belowcaptionskip}
\makeatother

\def\fnote#1#2{\begingroup\def\thefootnote{#1}\footnote{#2}
     \addtocounter{footnote}{-1}\endgroup}

\begin{document}
\title{Giant Leaps and Minimal Branes \\ in Multi-Dimensional Flux Landscapes}

\author{Adam~R.~Brown$^{1,2}$ \,and Alex~Dahlen$^{1}$ \vspace{.1 in}\\  
\vspace{-.3 em}  $^1$ \textit{\small{Physics Department, Princeton University, Princeton, NJ 08544, USA}}\\
\vspace{-.3 em}  $^2$ \textit{\small{Princeton Center for Theoretical Science, Princeton, NJ 08544, USA}}}
\date{}
\maketitle
\fnote{}{emails: \tt{adambro@princeton.edu, adahlen@princeton.edu}}

\begin{abstract}
\noindent There is a standard story about decay in multi-dimensional flux landscapes: that from any state, the fastest decay is to take a small step, discharging one flux unit at a time; that fluxes with the same coupling constant are interchangeable; and that states with $N$ units of a given flux have the same decay rate as those with $-N$. We show that this standard story is false. The fastest decay is a giant leap that discharges many different fluxes in unison; this decay is mediated by a `minimal' brane that wraps the internal manifold and exhibits behavior not visible in the effective theory.   We discuss the implications for the cosmological constant.

\end{abstract}

\section{Introduction} \label{sec:introduction}

Landscapes built of a large number of different fluxes give rise to a vast discretuum of vacua, and so will naturally contain ones that, like our own, have a tiny cosmological constant \cite{BP}.  In this paper, we show that transitions between these vacua typically occur by giant leaps, discharging many fluxes in unison; such transitions dramatically alter the cosmological constant. 

In a previous paper \cite{paper 1}, we found effects that enhance giant leaps, which relate to the presence of a radion. In this paper, we find a different set of enhancements, which exist even when the radion is fixed.  
In Sec.~\ref{sec:multiflux}, we study landscapes built of many genuinely different fluxes, and in Sec.~\ref{sec:monkey} we study the more sophisticated case where these different fluxes arise from a single higher-dimensional flux wrapping many different cycles of the internal manifold.

Our paradigmatic example is the Bousso-Polchinski (BP) landscape \cite{BP}.  Following BP, we treat the radion as fixed, and work in the thin probe-brane approximation.  Transitions in this landscape have received much attention \cite{bpprob}, but this paper is the first to identify the dominant  decay: giant leaps. We show that the branes that discharge different fluxes attract, so that decay proceeds by whole stacks of branes at once, rather than a single brane as was previously assumed. Though still generically exponentially suppressed, giant leaps are the fastest decay. In \cite{paper3}, we show that giant leaps persist away from these approximations by studying decays in an explicit radion stabilization, 6D Einstein-Maxwell theory with many distinct fluxes.\footnote{The possibility of nucleating stacks of branes in multi-dimensional flux landscapes is usually dismissed by citing \cite{GM}. That result, however, only applies to co-dimension one branes, for which gravity is repulsive---our branes are higher co-dimension, and gravity is attractive at short range. Furthermore, the results of \cite{GM} do not apply when the branes are charged under different fluxes.}

\section{Many fluxes} \label{sec:multiflux}

A collection of $\mathfrak{N}$ different fluxes, whose legs all point down the extra dimensions, makes a contribution to the cosmological constant, $\Lambda_\text{eff} = \Lambda_0 + \frac12 F^2$, where
\begin{equation}
\label{Fsquared}
F^2 = \sum_{i=1}^{\mathfrak{N}} g_i^2 N_i^2  \, ,
\end{equation}
$g_i$ is the charge of the $i$th magnetic quantum, and $N_i \in \mathbb{Z}$ is the number of units of the $i$th magnetic flux.   These fluxes may be discharged by nucleating a brane that forms a bubble in the extended directions.  If the brane carries $n_i$ units of charge, then inside the bubble $N_i \rightarrow N_i - n_i$ so that the energy density inside is reduced by
\begin{equation}
\frac{1}{2} \Delta F^2 = \frac{1}{2} \sum_{i=1}^{\mathfrak{N}} g_i^2 n_i(2N_i-n_i). \label{epsilon}
\end{equation}
The surface tension of the bubble is set by the brane tension; for the black branes, M-branes and D-branes we are interested in, that tension is
\begin{equation}
T \sim M_{4} \left( \sum_{i=1}^{\mathfrak{N}} g_i^2 n_i^2  \right)^{1/2} \label{tension},
\end{equation}
where $M_4$ is the 4D Planck mass. The rate to nucleate such a bubble is $\Gamma \sim e^{- B / \hbar}$, where in the thin-wall, probe-brane, semiclassical, no-4D-gravity approximation  \cite{Coleman}
\begin{equation}
B \sim \frac{T^4}{(\Delta F^2)^3}.
\label{rate}
\end{equation}

What is the fastest decay? It depends on the $g_i$s and the $N_i$s. For definiteness and simplicity we will focus throughout this paper on the case of identical fluxes $g_i = g$. We will also focus on a special class of states, the `diagonal' states that have $|N_i| = N$. These states are privileged because states off the diagonal, with unequal $|N_i|$, decay towards them; the largest $N_i$ discharge first, as these decays most rapidly reduce $F^2$. For diagonal states, one decay direction, `monoflux decays', is to drop $n$ units from a single flux, while leaving the other fluxes untouched. A second decay direction, `multiflux decays', is to drop just a single unit each from $n$ different fluxes. These definitions coincide for $n=1$, which has tension $T_1$ and rate exponent $B_1$. But as $n$ increases, there are two effects that make monoflux decays slower than multiflux decays.

{\bf $T$-effect}: For monoflux decays, $T = nT_1$.  This is linear because these extremal branes do not interact---their magnetic repulsion exactly cancels their gravitational attraction. For multiflux decays, $T = \sqrt{n} T_1$.  This is slower than linear because the branes bind together--- since they are charged under different fluxes, there is now no magnetic repulsion to cancel their gravitational attraction. Smaller $T$ means faster decay.

{\bf $\Delta F^2$-effect}: For monoflux decays, $\Delta F^2 =g^2 n (2N-n)$. This is slower than linear  because the flux lines repel. For multiflux decays, $\Delta F^2=g^2 n (2N-1)$.  This is linear because the flux lines are different and don't interact. Larger $\Delta F^2$ means faster decay.

The rate exponent $B$ can be evaluated for both decay directions.  Monoflux decays get slower with increasing $n$, $\, B_\text{mono} [n] > B_1$. But multiflux decays get faster
\begin{equation}
B_\text{multi}[n] = \frac{B_1}{n}. \label{eq:multirate}
\end{equation}
Indeed the fastest possible decay of a diagonal state is $n = \mathfrak{N}$, discharging one unit of each flux in unison and moving one step down the diagonal. (That moving many steps down the diagonal is slower follows from the monoflux result, with $g \rightarrow \sqrt{\mathfrak{N}} g$.)

If the $g_i$ vary, then the analog of the diagonal states are those with $|g_i N_i|$s as equal as possible consistent with quantization.  Decaying down the diagonal now may mean discharging more than a single unit of the fluxes with small $g_i$.
Gravitational corrections are easy to add; the form of the corrections was calculated by Coleman and De Luccia \cite{CDL}. While $B_1$ receives large gravitational corrections, Eq.~\ref{eq:multirate} is exact for decays from Minkowski. For decays from low and intermediate de Sitter,  it receives only small modifications, so that our parent vacuum is likely a giant leap away. From high de Sitter states, however, gravitational modifications are significant and small steps are the fastest decay. To confirm giant leaps away from the thin-wall, probe-brane approximation requires choosing an explicit flux compactification, as we have done in \cite{paper3}.

\section{Many cycles and minimal branes} \label{sec:monkey}

In the last section we saw that, in landscapes built of many genuinely different fluxes, the most likely decay is a giant leap. In this section we will investigate whether the same holds true when the `different' fluxes are really, from the higher-dimensional viewpoint, the \emph{same} flux, just wrapped around different compact cycles. 

Following BP \cite{BP}, let's compactify on a flat $m$-torus. To preserve macroscopic Lorentz symmetry, a $q$-form $\mathbf F$ that lives in this (3+1+$m$)-dimensional spacetime must point legs in \emph{no} macroscopic direction (and so $\star \mathbf F$ must point legs in \emph{every} macroscopic direction); after integrating out the extra dimensions, $\mathbf F$ becomes a scalar (and $\star \mathbf F$ a four-form).  

Let's consider an $\mathbf{F}$ that wraps a single basis $q$-cycle
 \begin{center}
\begin{tabular}{|l||c|c|c|c|c|c|c|c|c|c|} \hline
 {\color{white} \textrm{brane}}  & $\ t \ $& $\ r \ $ & $\ \theta \ $ & $\ \psi \ $  & $w_1$ & \ldots & $w_q$ &  \ldots & $w_m$ \\
\hline \hline
$\mathbf F $ \hspace{2.2mm}   &  &  &  & & x&  x&  x& &  \\
\hline
$\star \mathbf F $  & x & x& x & x & &  & & x & x\\
\hline
\end{tabular}
\end{center}\vspace{-.19in}
\makebox[5.35in][r]{.}\vspace{.1in}

\noindent This flux is quantized in units of $g/V_{q}$, where $g$ is the magnetic charge and $V_q$ is the $q$-volume of the wrapped cycle. After integrating out all the extra dimensions, we are left with a term 
\begin{equation}
\frac12 V_m F^2=\frac12 \frac{V_{m-q}}{V_{q}} g^2 N^2
\end{equation}
in the Lagrangian, where $V_m=V_qV_{m-q} = L_1 L_2 \ldots L_m$ is the total internal volume, so that the effective 4D  magnetic charge is $g_i=g\sqrt{V_{m-q}/V_q}$. This flux is discharged by a $(2+m-q)$-brane that couples magnetically to $\mathbf{F}$ and forms a bubble in the extended dimensions.  The remaining $m-q$ legs wrap the dual cycle 
 \begin{center}
\begin{tabular}{|l||c|c|c|c|c|c|c|c|c|c|}
 \hline {\color{white} brane}  & $\ t \ $& $\ r \ $ & $\ \theta \ $ & $\ \psi \ $  & $w_1$ & \ldots & $w_q$ &  \ldots & $w_m$ \\ \hline
\hline
\textrm{brane}  & x & {\color{white} x} & x & x & {\color{white} x} & {\color{white} x} &  {\color{white} x} &  x&  x  \\
\hline
\end{tabular}
\end{center}\vspace{-.17in}
\makebox[5.35in][r]{.}\vspace{.08in}

\noindent To see why the dual cycle is appropriate, consider integrating the magnetic source equation $\mathbf{dF=\star j}$ in a Gaussian pillbox across the brane.  Restricting to the extra dimensions, if this flat brane is spanned by one forms $\{ {\mathbf{Y_{1}}}, \ldots , {\mathbf{Y_{m-q}}} \}$, then
\begin{equation}
\label{decomposable}
\Delta \mathbf F=\star \frac{g}{V_m} \mathbf Y_{1} \wedge\dots\wedge \mathbf Y_{m-q},
\end{equation}
and indeed this works for $\mathbf{Y}_{1} = L_{q+1} dw_{q+1}$, $\mathbf{Y}_{2} = L_{q+2} dw_{q+2}$, etc.   
The effective tension of this brane is proportional to the fundamental tension times its volume in the extra dimensions,
\begin{equation}
T\sim \left(g M_{4+m}^{\, 1 + \frac{m}{2}}\right)V_{m-q}, 
\end{equation}
where $M_{4+m}^{\, 2 + m} = M_4^{\, 2} / V_m$; again this is consistent with $g_i = g \sqrt{V_{m-q}/V_{q}}$.

A general $\mathbf{F}$ will have components along many of the $\mathfrak{N} = {m \choose q}$ basis $q$-cycles of the $m$-torus
 \begin{equation}
 \mathbf{F}=\sum_{\substack{\text{permutations}\\i=1}}^\mathfrak{N}\frac{ g_i N_i }{\sqrt{V_m}}\;{d}w_{p_1(i)} \wedge \dots\wedge {d}w_{p_q(i)}.
 \end{equation}
The cross-terms in $F^2$ all vanish so we recover Eq.~\ref{Fsquared}, and with it the $\Delta F^2$-effect.  What of Eq.~\ref{tension} and the $T$-effect?

To discharge many fluxes at once, the brane must wrap many basis co-cycles---the appropriate configuration will be the one that does this with minimum possible area, and so minimum possible effective tension.  In the remainder of this section, we study this constrained optimization problem; we start with a few simple examples and build up to the general case.  For now, we set $g$ and the sides of the torus to 1.

For a $\mathbf{q=1}${\bf -form} on an $\mathbf{m=2}${\bf -torus}, $\mathfrak N= {2 \choose 1} = 2$.  To discharge a flux wound round the $w_1$-direction ($\Delta \mathbf F=dw_1$) requires a brane wound round the $w_2$-direction ($\mathbf Y= -dw_2$), and vice versa.  To discharge both at once, $\Delta \mathbf F=dw_1+dw_2$, one possibility is to use two different branes, wrapping the $w_2$- and $w_1$-directions respectively, for a total tension of $2$.  But you can do better.  The brane can instead stretch across the diagonal ($\mathbf Y=dw_1-dw_2$), so that it still wraps the same cycle but has lower area, for a tension of $\sqrt 2$:
 \begin{figure}[h!] 
  \centering
  \includegraphics[width=4.3in]{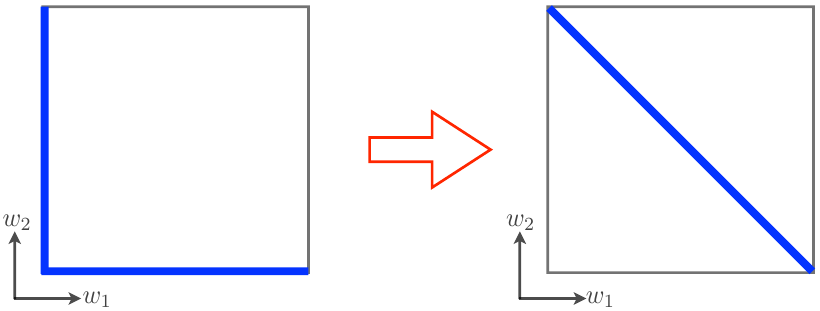}
  \label{fig:5d1each}\vspace{-.4in}
\makebox[4.8in][r]{.}
\end{figure}
\vspace{.08in}

\noindent This diagonal brane is the shortest brane that achieves this transition\footnote{Hashimoto and Taylor \cite{Taylor} showed that D1-branes that intersect as in the left pane have a tachyon in their spectrum, which represents the instability of the branes to reconnect and form the bound state configuration in the right pane.}  (it's the minimum area element of its homology class, the `minimal brane'), and the $T$-effect is retained.
\pagebreak

Analogous configurations can be found for any combination of fluxes dropped. To discharge $\Delta \mathbf F= n_1dw_1+n_2dw_2$, the minimal brane is stretched in the normal direction to the flux dropped, $\mathbf Y=n_2dw_1-n_1dw_2$, and has length $\sqrt{n_1^{\,2}+n_2^{\,2}}$.  For example, 

\begin{figure}[h!] 
  \centering
 \includegraphics[width=5.5in]{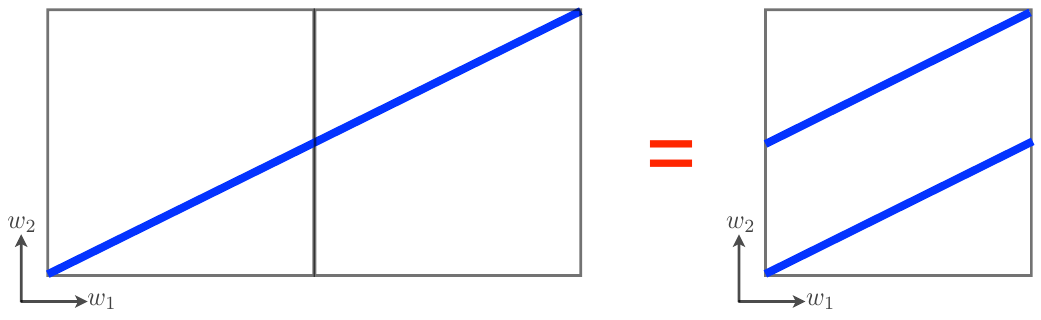}
  \label{fig:5dMultiEach} \vspace{-.1in}
\end{figure}

\noindent is the minimal brane with $n_1=-1$ and $n_2=2$, shown both in the covering space and projected onto the unit cell. In this case, Eq.~\ref{tension} just amounts to Pythagoras' Theorem.


For a $\mathbf{q=1}${\bf -form} on an  $\mathbf{m=3}${\bf -torus}, $\mathfrak N= {3 \choose 1} =3$. Discharging $\Delta \mathbf F=dw_1$ takes a two-brane stretched along the $w_2$-$w_3$ cycle, and so on.  As before, discharging $\Delta \mathbf{F}=dw_1+ dw_2$ doesn't take two branes and total tension 2, but only a single flat brane with $\mathbf{Y_1} \wedge \mathbf{Y_2}=(dw_1-dw_2)\wedge dw_3$ and total tension $\sqrt 2$:
\begin{figure}[h!] 
  \centering
 \includegraphics[width=4in]{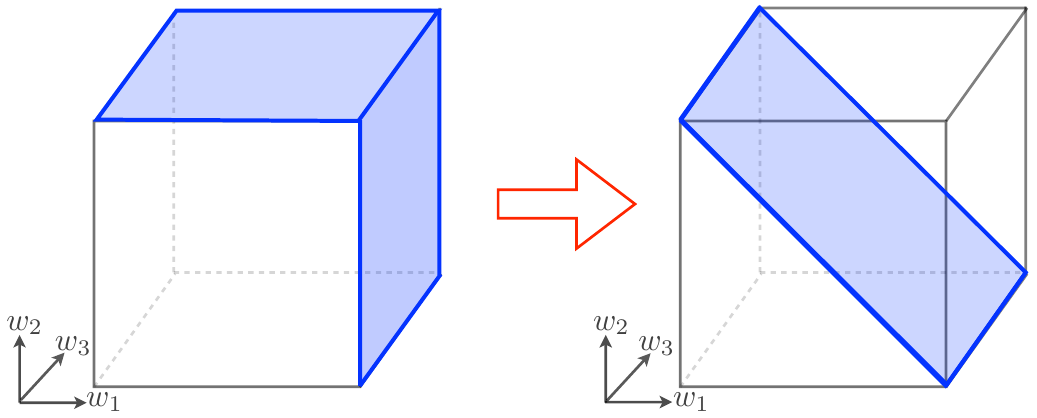}
  \label{fig:6dNeighbours} \vspace{-.32in}
\makebox[4.4in][r]{.}
\end{figure}
\vspace{.12in}

\noindent Indeed, for any $\Delta \mathbf{F}=n_1 dw_1 + n_2 dw_2 + n_3 dw_3$ there is a minimal brane with tension $\sqrt{n_1^{\,2}+n_2^{\,2}+n_3^{\,2}}$; the normal to the brane points in the direction of the dropped flux.  For example, 

 \begin{figure}[h!] 
  \centering
  \includegraphics[width=6in]{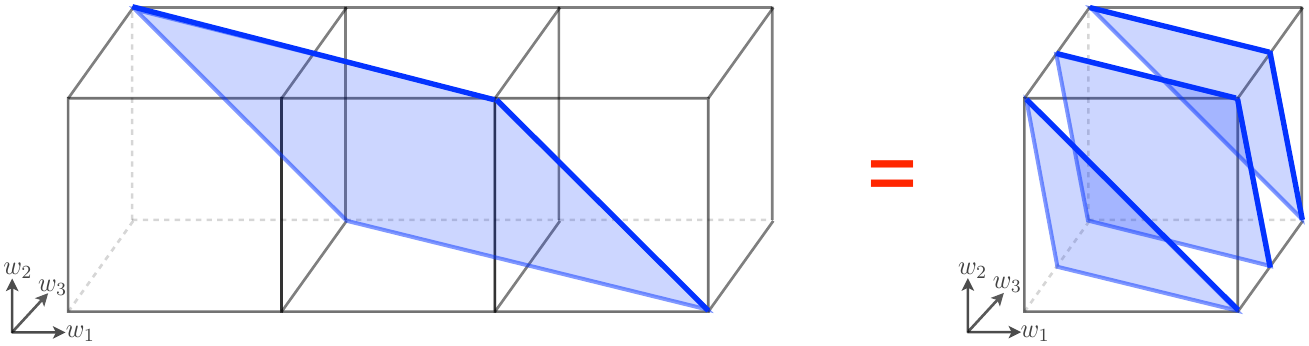}
  \label{fig:6dDouble} \vspace{-.1in}
\end{figure}

\noindent is the minimal brane that drops $n_1=1, n_2=1,n_3=2$.

Up to now it has all been straightforward---we could unwrap any set of fluxes in any combination with a single flat brane, and the story has been the same as it was for the genuinely different fluxes of Sec.~2. In higher dimensions, however, this is no longer true. A $\mathbf{q=2}${\bf -form} on an $\mathbf{m=4}${\bf -torus}, which has $\mathfrak N = {4 \choose 2} =6$, has all the complexities of the general case, and we will use it to exemplify them. 

The source of these complexities is that you cannot always discharge a given flux with a single flat brane. For example, to discharge
\begin{equation}
\Delta \mathbf{F} = d{w}_1 \wedge d{w}_2 + d{w}_3 \wedge d{w}_4
\label{uhoh}
\end{equation}
takes two branes, one that spans the $w_3$-$w_4$ plane and another that spans the  $w_1$-$w_2$ plane. There is no better way: these two branes only intersect at a point and cannot reduce their area by coalescing.\footnote{Another way to see this is that for D2-branes, this configuration is supersymmetric \cite{Johnson}.} The problem is that this $\Delta \mathbf{F}$ cannot be written in the form $\mathbf{Y}_1 \wedge \mathbf{Y}_2$. If it could, then by Eq.~\ref{decomposable}, these $\mathbf{Y}$s span a single flat brane that would discharge this $\Delta \mathbf F$; the right-hand side of Eq.~\ref{decomposable} is the volume form of the brane, so its effective tension would be given by Eq.~\ref{tension} and we would recover the full $T$-effect.   The property that $\Delta \mathbf F$ is writeable in this form is known as decomposability, and a two-form is decomposable if and only if it has just a single pair of nonzero eigenvalues.  [One-forms and $(m-1)$-forms are always decomposable, which is why this behavior first appears at $m=4$.]

A direct consequence is that the different effective 4D fluxes are not now interchangeable.  In Sec.~2, when any two fluxes discharged together, the brane had tension $\sqrt2$.  Here, it depends \emph{which} two fluxes.  Neighboring fluxes can be discharged together with tension $\sqrt2$ [e.g.~$\Delta\mathbf F=dw_1\wedge(dw_2+dw_3)$ is decomposable] whereas the two fluxes of Eq.~\ref{uhoh} require tension 2.  The rotational symmetry of the landscape is broken---switching a pair of occupation numbers, $N_i \leftrightarrow N_j$, can change the decay rate.

Bizarrer still, the reflection symmetry is broken too---switching the sign of one occupation number, $N_i\rightarrow-N_i$ (and so $n_i\rightarrow-n_i$), can also change the decay rate. For instance, $\Delta \mathbf F=(dw_1+dw_2)\wedge(dw_3+dw_4)$ is decomposable and discharges fast, whereas the same combination with the sign of the $dw_1\wedge dw_3$ term flipped is not decomposable and discharges slowly.

These complexities mean that finding the fastest decay is a more intricate problem than it was for the many genuinely different fluxes of Sec.~2.  Let's start, as before, by considering decay from a diagonal state with all the occupation numbers equal $|N_i|=N$ (and continue using $m=4$, $q=2$ as our prototype). The fastest decay need no longer be to discharge one unit of every type of flux.  Such a $\Delta \mathbf F$ is not decomposable, so the tension must be larger than $\sqrt6$, and some of the $T$-effect is lost.  But some of the $T$-effect is preserved: we can certainly do better than 6 separate branes by clustering neighbors and allowing them to coalesce. But even this is not the ideal configuration---these diagonal branes still intersect along a line and want to pull tight. The resulting optimal configuration cannot be flat, since $\Delta \mathbf F$ is not decomposable; indeed the minimal brane is generally curved \cite{MinimalSurfaces}.  

Because of this impediment, decays that would have been subdominant in Sec.~2 can now be fastest, so that diagonal states need not decay to diagonal states.  For example, it may be best to discharge fewer than $\mathfrak{N}$ fluxes just so that the minimal brane can be flat and the full $T$-effect retained. If the signs are right, five fluxes may be discharged in unison, $\Delta \mathbf F=(dw_1+dw_2+dw_3)\wedge(dw_3+dw_4)$, by a single flat brane. On the other hand, it may be best to discharge \emph{more} flux. One unit each of five fluxes and two units of the sixth can be discharged in unison, $\Delta \mathbf F=(dw_1+dw_2-dw_3)\wedge(dw_2+dw_3+dw_4)$, by a single flat brane. This can be the fastest decay, so that the leaps can be even gianter than in Sec.~\ref{sec:multiflux}.

For higher dimensional tori, the story follows analogously.  [One case of special interest is the BP case, of M-theory on a flat $m=7$-torus with a $q=4$-form flux, so $\mathfrak N = {7 \choose 4} = 35$. There, a flat brane can discharge a single unit each of at most 21 of the 35 fluxes, $\Delta \mathbf{F}=(dw_1+dw_2+dw_3)\wedge(dw_3+dw_4+dw_5)\wedge(dw_5+dw_6+dw_7)$.]  As before, states that are off the diagonal will decay towards it, and if the $g_i$ vary, the fastest decays will stick to the approximate diagonal.  

\section{Conclusion}

The innovation of the Bousso-Polchinski model \cite{BP} is to introduce many different kinds of flux, giving a huge discretuum of vacua including many in the habitable zone. We have shown that, precisely because of the many fluxes, decay proceeds by giant leaps.  Giant leaps are mediated by minimal branes which, already for tori, exhibit exotic behavior not visible in the effective theory.  We expect more general manifolds to have yet richer decay phenomenology, with consequences for the stability of stringy de Sitter, probabilities in the landscape \cite{bpprob}, and the cosmological constant \cite{staggering}.

\section*{Acknowledgements}
Thanks to Raphael Bousso, Steve Gubser, Chris Herzog, Igor Klebanov, Silviu Pufu, Paul Steinhardt, Herman Verlinde and Masahito Yamazaki for helpful discussions.


\begin{thebibliography}{99}
  
\bibitem{BP}
  R.~Bousso and J.~Polchinski,
  ``Quantization of four-form fluxes and dynamical neutralization of the
   cosmological constant,''
  JHEP {\bf 0006}, 006 (2000)
  [arXiv:hep-th/0004134].


\bibitem{paper1}
 A.~R.~Brown and A.~Dahlen,
 ``Small Steps and Giant Leaps in the Landscape,''
 Phys.\ Rev.\ D {\bf 82}, 083519 (2010)
  [arXiv:1004.3994 [hep-th]].

  \bibitem{bpprob}  
    T.~Clifton, S.~Shenker and N.~Sivanandam,
  ``Volume Weighted Measures of Eternal Inflation in the Bousso-Polchinski
  Landscape,''
  JHEP {\bf 0709}, 034 (2007)
  [arXiv:0706.3201 [hep-th]];
  
    R.~Bousso and I.~S.~Yang,
  ``Landscape Predictions from Cosmological Vacuum Selection,''
  Phys.\ Rev.\  D {\bf 75}, 123520 (2007)
  [arXiv:hep-th/0703206];
  
    D.~Schwartz-Perlov,
  ``Anthropic prediction for a large multi-jump landscape,''
  JCAP {\bf 0810}, 009 (2008)
  [arXiv:0805.3549 [hep-th]].
 


\bibitem{paper3}
  A.~R.~Brown and A.~Dahlen,
  ``Bubbles of Nothing and the Fastest Decay in the Landscape,''
  Phys.\ Rev.\ D {\bf 84}, 043518 (2011)
  [arXiv:1010.5240 [hep-th]].
  
  \bibitem{GM}
  J.~Garriga and A.~Megevand,
  ``Coincident brane nucleation and the neutralization of Lambda,''
  Phys.\ Rev.\  D {\bf 69}, 083510 (2004)

 \bibitem{Coleman}
  S.~R.~Coleman,
  ``The Fate Of The False Vacuum. 1. Semiclassical Theory,''
  Phys.\ Rev.\  D {\bf 15}, 2929 (1977);
  
  J.~D.~Brown and C.~Teitelboim,
  ``Neutralization of the Cosmological Constant by Membrane Creation,''
  Nucl.\ Phys.\  B {\bf 297}, 787 (1988).
  
 \bibitem{CDL}
  S.~R.~Coleman and F.~De Luccia,
  Phys.\ Rev.\  D {\bf 21}, 3305 (1980).

 \bibitem{Taylor}
 A.~Hashimoto and W.~Taylor,
 ``Fluctuation spectra of tilted and intersecting D-branes from the
 Born-Infeld action,''
 Nucl.\ Phys.\  B {\bf 503} (1997) 193

\bibitem{Johnson}
 C.~V.~Johnson,
 ``D-Branes,''
Chapter 11, 
{\it  Cambridge University Press} (2003).

 \bibitem{MinimalSurfaces}
 R.~Osserman,
``Survey of Minimal Surfaces,''
 {\it Van Nostrand Reinhold Press} (1969).

 \bibitem{staggering}
  D.~Schwartz-Perlov and A.~Vilenkin,
  ``Probabilities in the Bousso-Polchinski multiverse,''
  JCAP {\bf 0606}, 010 (2006)
  [arXiv:hep-th/0601162].


\end{thebibliography}
\end{document}